\documentclass[lettersize,journal]{IEEEtran}
\usepackage{amsmath,amsfonts}
\usepackage{physics,amsmath}
\usepackage{algorithm}
\usepackage{algpseudocode}
\usepackage{array}
\usepackage{bm}
\usepackage{subcaption}

\usepackage{mathtools}
\usepackage{xcolor}
\usepackage{subcaption}
\usepackage{textcomp}
\usepackage{stfloats}
\usepackage{lipsum}  
\usepackage{url}
\usepackage{verbatim}
\usepackage{graphicx}
\usepackage{cite}
\usepackage{pdfpages}
\usepackage{tikz}
\usetikzlibrary{3d}
\usetikzlibrary{arrows.meta} 

\DeclareCaptionLabelFormat{simple}{#2}
\begin{document}

\title{
3D Receiver for Molecular Communications  in Internet of Organoids}
\author{Shaojie Zhang,~\IEEEmembership{Student             Member,~IEEE}
        and Ozgur B. Akan,~\IEEEmembership{Fellow,~IEEE}
        \thanks{Shaojie Zhang is with the Internet of Everything Group, Electrical
        Engineering Division, Department of Engineering, University of Cambridge,
        CB3 0FA Cambridge, U.K. (e-mail: sz466@cam.ac.uk). }
        \thanks{Ozgur B. Akan is with the Internet of Everything Group, Electrical
        Engineering Division, Department of Engineering, University of Cambridge,
        CB3 0FA Cambridge, U.K., and also with the Center for neXt-Generation
        Communications (CXC), Department of Electrical and Electronics Engineering, Koc¸ University, 34450 Istanbul, Turkey (e-mail: oba21@cam.ac.uk;
        akan@ku.edu.tr)}
        \thanks{This work was supported in part by the AXA Research Fund (AXA Chair
for Internet of Everything at Koc¸ University).}
}


\maketitle

\begin{abstract}

Organoids have garnered attention due to their effectiveness in modeling the 3D structure of organ interactions. However, the communication engineering perspective has received relatively little attention. One way to achieve organoids communication is molecular communication (MC). Molecular communication is a bio-inspired communication paradigm that uses molecules as information carriers. It is considered one of the most promising methods for enabling the Internet of Nano-Things (IoNT) and nanonetworks. BioFETs are commonly used to implement practical MC receivers. However, most previous analyses have focused on a planar device, neglecting considerations like the threshold voltage and its potential 3D structure. This paper introduces the first FinFET-based MC receiver that covers both the top and side gates with receptors. Both binding noise and flicker noise are considered in the analysis. The performance, in terms of signal-to-noise ratio (SNR) and symbol error probability (SEP), is compared with that of the 2D receiver.

\end{abstract}


\section{Introduction}

\IEEEPARstart{O}{rganoids} are fundamental in vitro, tissue-engineered cell models derived from self-organizing human stem cells. They mimic many intricate details of the structure and function of corresponding in vivo tissues\cite{Zhao2022organoids}. Due to their advantages over the more expensive and less reliable traditional animal models, organoids are increasingly utilized in biomedical research. Unlike 2D tissue cultures, which cannot accurately model drug diffusion kinetics and dosage effectiveness, organoids successfully simulate three-dimensional tissue architectures and the physiological fluid flow conditions essential for maintaining normal tissue environments. This makes them a more effective tool for studying drug interactions and effectiveness\cite{Skardal2016organoidreview},\cite{Marx2012Humanonachip}.

A significant barrier to advancing biological complexity and in vivo-like functionality in the early stages of organoids cultivation is the absence of a perfusable vasculature\cite{Zhang2021vascularizedorganoids}. Microfluidic devices replace the vasculature and enable controlled perfusion of oxygen, nutrients, and growth stimulants as well as the removal of waste products, permitting a more physiologic-like differentiation in the direction of a more intricate, advanced, and 'in vivo-like' model\cite{Smirnova2023OI}.

In most cases, a single organoid cannot fully replicate the drug effectiveness in human body since it does not take into account the interaction between multiple tissues. For example, in cancer metastasis, multiple tissue sites and circulatory systems are involved, which underscore the significance of multi-organoids system\cite{Skardal2016organoidreview}. Multi-organoid systems have already been implemented in the past, for instance, in \cite{Miller201614compartment}, authors proposed a microfluidic cell culture device representing 13 organs to model the inter-organ crosstalk and to assess the relationship between organ volume and blood residence time. The whole cascade of cancer metastasis, including the extravasation of cancer cells from the tumor, their movement through the bloodstream, and their penetration of an external organ are modelled in \cite{Lai2017vasculaturedynamic}. Comparatively, little attention has been given to the communication engineering perspective of organoid interactions. Therefore, the Internet of Organoids aims to provide seamless communication for organoids to organoids, organoids to electronic entities, and organoids to organs. More specifically, the Internet of Organoids involves viewing this biological system through the lens of communication system design, data transmission, and signal processing. 

One common way that human tissues interact with each other is molecular communication (MC). MC is a biologically inspired technology that encodes, transmits, and receives information through the use of messages conveyed in patterns of molecules\cite{Akyildiz2015internetofbionanothings},\cite{Nakano2013MCbook}. Over billions of years, living things have already chosen and utilized fully functional molecular communication (MC) networks through evolutionary processes. This method is proven to be bio-compatible, and require very little energy to generate and propagate. As a result, it is regarded as one of the most promising methods for enabling Internet of Nano-Things (IoNT) and nanonetworks\cite{Akan2016fundamentalsofMC},\cite{Nakano2012MCnetworkingreview}. 

MC has been investigated from various perspectives. Many studies focus on the modulation scheme, detection methods, and information-theoretical models of MC channels. However, most of these studies ignore the physical properties of receiver signals within the communication channel \cite{Kuscu2016MCreceiver}. In the Internet of Nano-Things, nanomachines are considered as potential receivers. Bio-transceiver and bio-nanomachines have been proposed in \cite{Nakano2014bionanomachine} and \cite{Unluturk2015bacteriareciver}, respectively. While creating nanomachines only from biocomponents has the benefit of biocompatibility, there are some drawbacks that limit the range of applications that nanonetworks can be used for. The low computational capabilities of biocomponents significantly confine the implementation of complex communication protocols and algorithms\cite{Unluturk2015bacteriareciver}. At same time, they can only operate \textit{in vivo} application. Moreover, biocomponents create difficulties in seamlessly connecting organoids to electronic entities such as computers, thus challenging to integrate organoids to the Internet of Nano-Things (IoNT)\cite{Akyildiz2015internetofbionanothings}. Contrarily, artificial MC receiver can conduct \textit{in situ}, continuous, label-free operation for both \textit{in vivo} and \textit{in vitro} applications\cite{Kuscu2019surveyontranceiver},\cite{Kuscu2016MCreceiverbiosensor}. An artifical MC reciver will selectively detect the targetted information ligands concentration and transfer to a more understandable signal. 

Thanks to new nanomaterials such as nanowires, carbon nanotubes (CNTs), and graphene, FET-based biosensors, or bioFETs, meet the primary requirements of an MC receiver with enhanced performance. For example, a silicon nanowire (SiNW) bioFET-based MC receiver is modeled in \cite{Kuscu2015modelingbiofet} and \cite{Kuscu2016MCreceiverbiosensor}, which discuss the structures of bioFET-based MC receivers. These investigations were expanded in \cite{Kuscu2016SiNWmodeling}, which includes a thorough noise analysis based on the equilibrium assumption for the receptor-ligand response at the receiver surface. However, there are still some unresolved questions in the modeling. In \cite{Kuscu2016SiNWmodeling}, the authors suggest an ideal BioFET with a 0V threshold voltage so that the device operates in the triode region during detection. However, achieving a 0V threshold voltage is challenging in real life. High gate voltage could potentially cause difficulty in power supply.

In system modeling, the MC receiver is usually considered as a spherical entity\cite{Pierobon2011diffusionbasedMC}. However, in previous work \cite{Kuscu2016SiNWmodeling}, \cite{Kuscu2016MCreceiver}, although a gate-all-around structure has already been proposed, the full analysis is based on 2D MOSFETs, and therefore, cannot demonstrate its 3D properties. FinFET, due to its improved subthreshold slope (SS), better stability, higher (ION/IOFF) ratio, enhanced short channel performance, and smaller intrinsic gate capacitance\cite{Taur1998book},\cite{Chen1987MOSFET}, has become the ideal candidate for a 3D MC receiver. Up until now, many papers have explored new FET-based structures. For example, nanowire FET-based biosensors are proposed in \cite{Ahn2012nanowirefet}, and junctionless gate-all-around nanowire field-effect transistors with an extended gate for biomolecule detection are presented in \cite{Chen2019gaafet}. However, none of these have performed the analysis from a communication engineering perspective. For the first time, the signal-to-noise ratio (SNR), and symbol error probability (SEP) of a 3D MC receiver are carried out in the work, taking into account the significance of signaling particles in MC transceiver design.

The rest of the paper is organised as follows. In Section \ref{3D BioFET Working Principle And Modeling}, we describe our proposed 3D BioFETs and explain its operation principle. We develop the model of SiNW FET-based MC receivers in Section \ref{SiNW FinFET Receiver Model}. Section \ref{Performance Analysis}  includes an evaluation of noise, signal-to-noise ratio (SNR), and symbol error probability (SEP).
These aspects will be presented and contrasted with those of
the planar design. Finally, our key findings are concluded in Section \ref{Conclusion}.

\color{black}

\section{3D BioFET Working Principle And Modeling}
\label{3D BioFET Working Principle And Modeling}

\begin{figure}[t]
    \centering
    \begin{subfigure}[b]{0.45\textwidth}
        \centering
        \includegraphics[width=\textwidth]{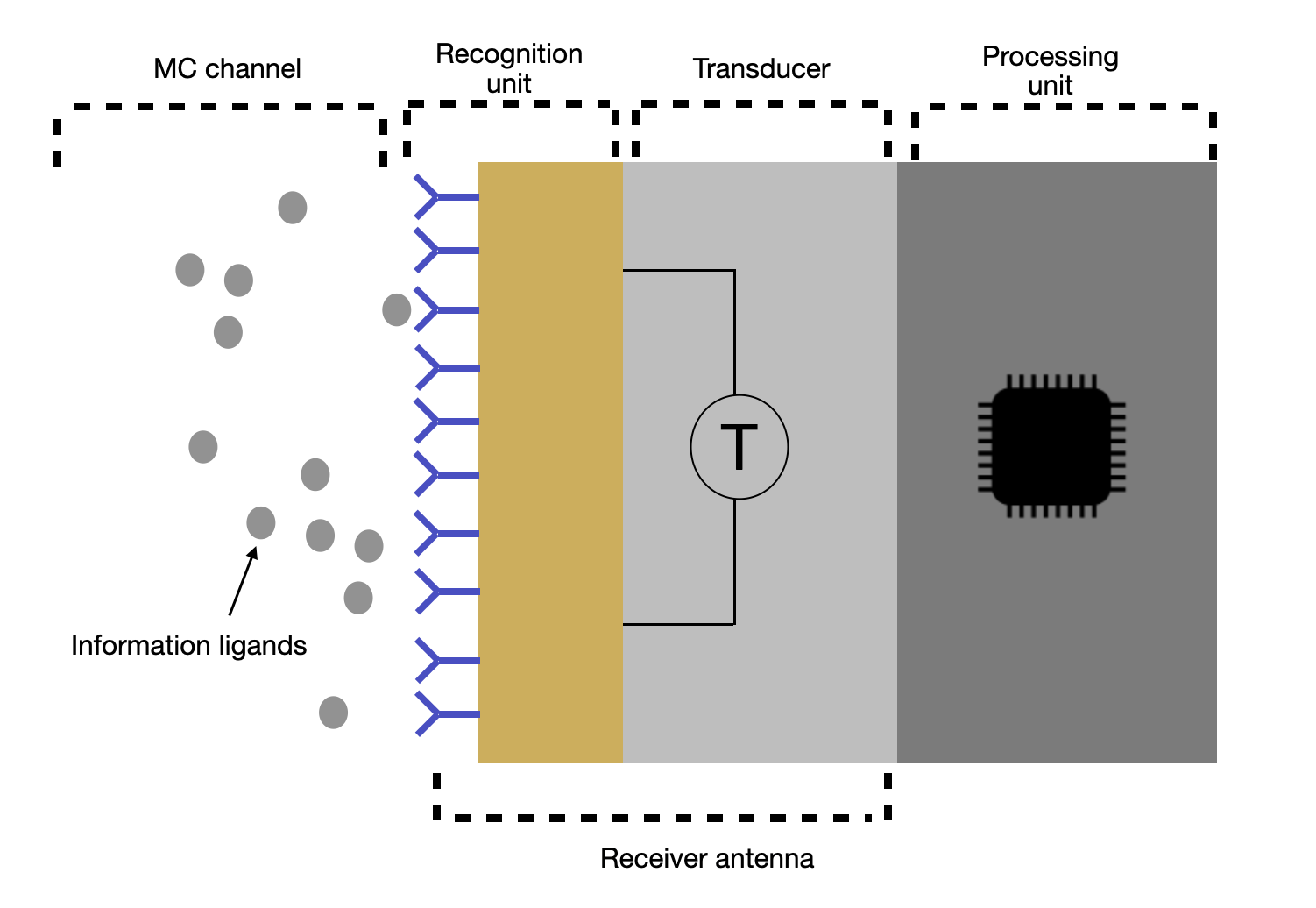}
        \caption{}
        \label{fig:mcreceiver-functional-units}
    \end{subfigure}
    \hfill 
    \begin{subfigure}[b]{0.45\textwidth}
        \centering
        \includegraphics[width=\textwidth]{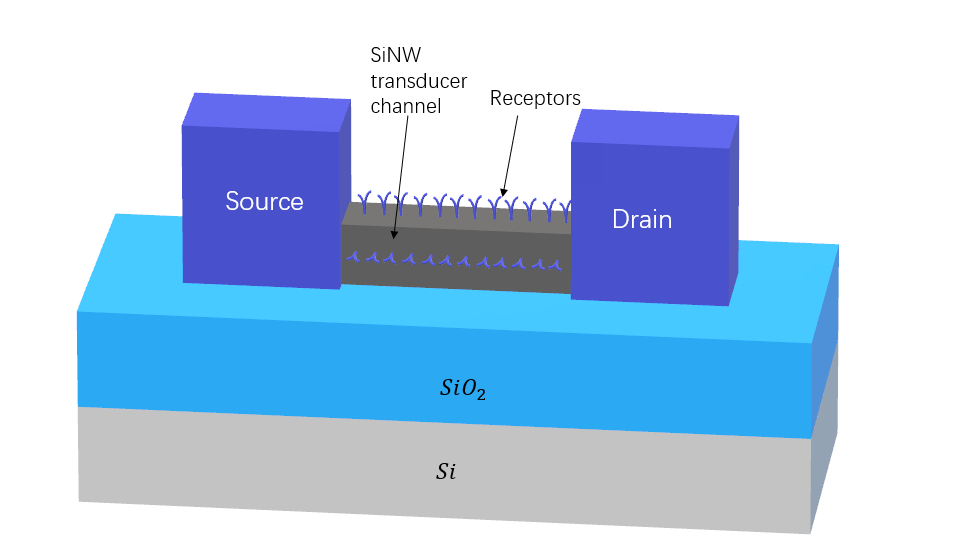}
        \caption{}
        \label{fig:finfet-concepture-receiver}
    \end{subfigure}
    \caption{(a) Functional units of an MC receiver \cite{Kuscu2015modelingbiofet}, and (b) 3D
    SiNW FET-based MC receiver antenna where receptors cover all the gate area. Insulating $SiO_2$ layer entirely covering the SiNW, source and drain is not shown for better visualization of the transducer}
    \label{fig:finfet-concepture-receiver-and-mcreceiver-functional-units}
\end{figure}
A BioFET consists of a recognition unit, a transducer, and a processing unit, as illustrated in Figure \ref{fig:mcreceiver-functional-units}. Information molecules propagate in the MC channel and approach the recognition unit. When ligands reach the gate area, receptors identify and bind with the appropriate ligands. The transducer then converts this signal, originally encoded in the concentration of information molecules, into an electrical potential\cite{Kuscu2015modelingbiofet}. The potential generated at the gate is proportional to the accumulated ligands. Upon applying a potential difference to the drain and source, a current flows through them. This gate potential modulates the channel's conductance, resulting in varying current levels. Figure \ref{fig:finfet-concepture-receiver} depicts the conceptual 3D FinFET MC receiver we propose, where all three sides of the SiNW channel are covered with receptors.

\begin{figure}[t]
    \centering
    \begin{subfigure}[b]{0.45\textwidth}
        \includegraphics[width=\textwidth]{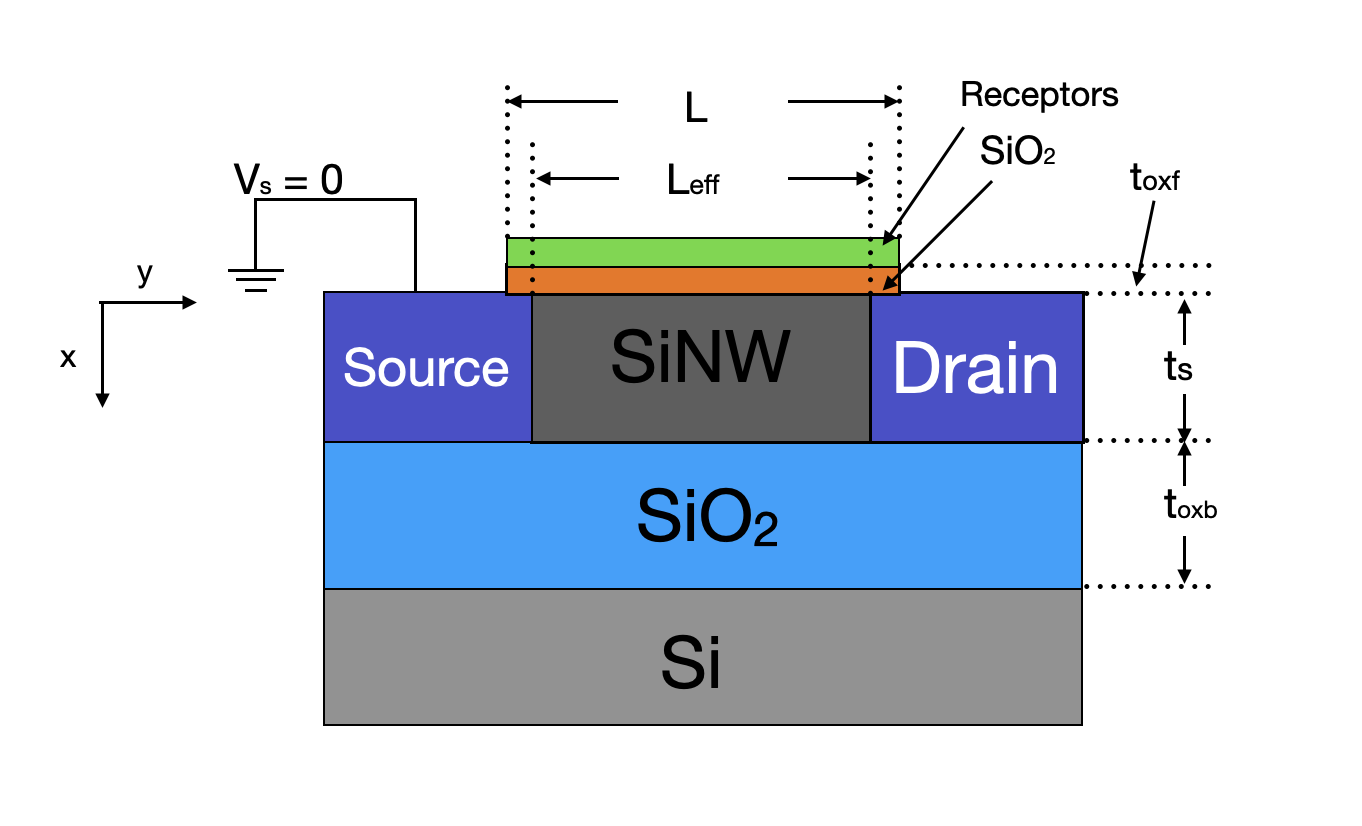}
        \caption{}
        \label{Cross sectional view along the channel length}
    \end{subfigure}
    \hfill
    \begin{subfigure}[b]{0.45\textwidth}
        \includegraphics[width=\textwidth]{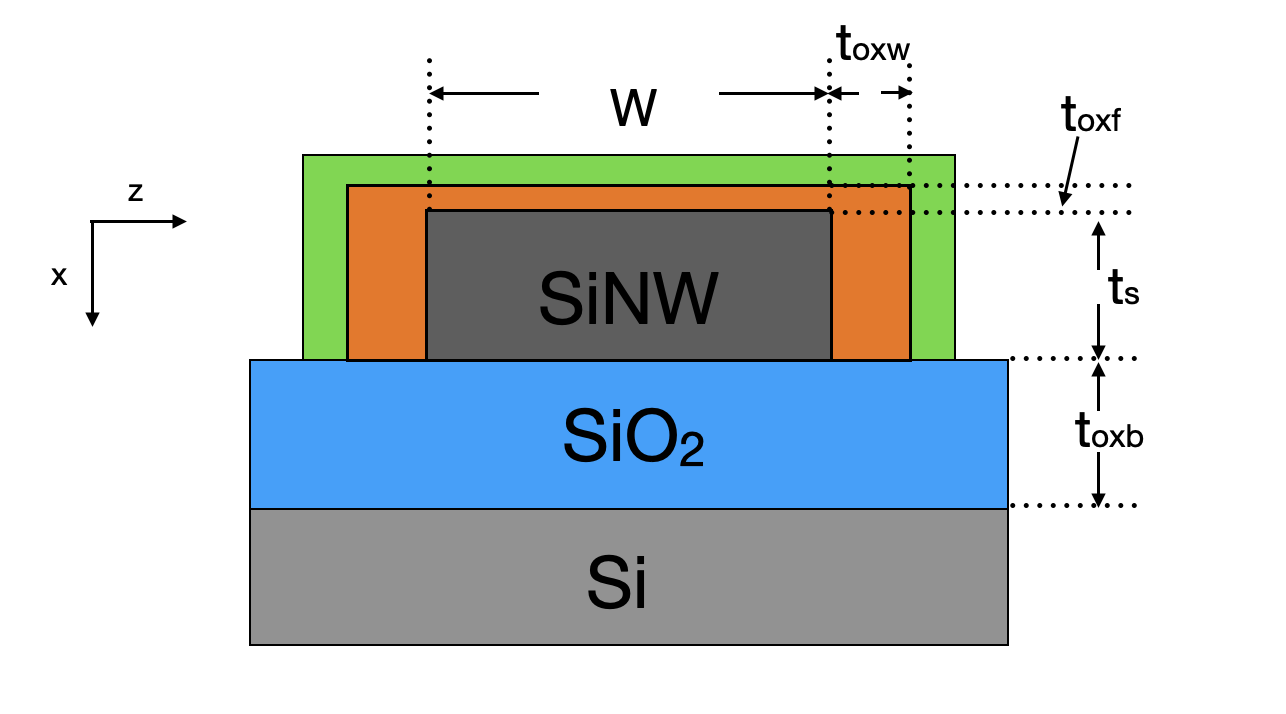}
        \caption{}
        \label{Cross sectional view along the channel width}
    \end{subfigure}
    \caption{Cross-sectional views: (a) along the channel length and (b) along the channel width.}
    \label{fig:cross_sectional_view}
\end{figure}

For the FinFET-based receiver, due to its unique 3D structure, using the original MOSFET (2D) based current equation and transconductance equation will no longer be an accurate estimation. FinFET improves the control of the gates over the MOSFET channel as the gate voltage is applied from the top and sides. To better understand the structure, Figure \ref{Cross sectional view along the channel length} shows a cross-sectional view of the proposed device along the channel length. To more accurately quantify the location within the channel, we define the source and the drain that are at $y = 0$ and $ y = y_{eff} $, respectively, where $y_{eff}$ is the effective channel length. The front and back interfaces between $SiNW-SiO_2$ are defined as $x = 0$ and $ x = t_s$ where $t_s$ represents the thickness of SiNW. The front and back oxide thicknesses are $t_{oxf}$ and $t_{oxb}$. The cross-sectional view along the channel width is shown in Figure \ref{Cross sectional view along the channel width}. Two $Si-SiO_2$ interfaces are located in $z=0$ and $z=W$. $t_{oxw}$ is the side wall oxide thickness. In our model, we assume the source and the body are perfectly grounded. For simplicity, we assume $t_{oxf} = t_{oxw} = t_{ox}$. Therefore, the area of functionalized surface, i.e., $A_r$ can be approximated as 
\begin{equation}
    A_r = (W + t_s \times 2) \times L_{eff} 
\end{equation}
where $L_{eff}$ is the effective length of the channel in y direction. Assume the width of oxide layer is negligible compared to the height of the channel. 

The threshold voltage, $V_t$, is a key parameter for FET-based transistors. In many previous works on modeling FET-based receivers, the threshold voltage has not been taken into account\cite{Kuscu2015modelingbiofet}, \cite{Aktas2022WSK}. The threshold voltage of a long-channel n-type FinFET is expressed as
\begin{equation}
    V_t=V_{f b}-\frac{2 k T}{q} \ln \left(\frac{q t_{ox}}{\varepsilon_{o x}} \sqrt{\frac{n_i^2 \varepsilon_{SiNW}}{2 k T N_A}}\right),
\end{equation}
where $V_{fb}$ is the flat-band voltage, $k$ is the Boltzmann constant, $T$ is the temperature in Kelvin. $n_i$ is the  intrinsic carrier concentration and $N_A$ is the doping concentration. q is the charge of an electron, $\varepsilon_{SiNW}$ and $\varepsilon_{ox}$ are the permittivity of SiNW and oxide, respectively\cite{Tsormpatzoglou2012FinFET}. 

For most tri-gate FinFET (similar structure as in Figure \ref{fig:finfet-concepture-receiver}), $t_s > W $. With this simplification, most of the current flows along the side gates. Therefore, the effective width of the device becomes $ W_{eff} =W + t_s \times 2 $, and hence, the drain current $I_d$, in FinFET can be found by
\begin{equation}
    I_d=\mu \frac{2 W_{eff}}{L_{eff}} \frac{\varepsilon_{ox}}{t_{oxf}}\left(\frac{2 k T}{q}\right)^2\left[\left(q_{i s}-q_{i d}\right)+\frac{1}{2}\left(q_{i s}^2-q_{i d}^2\right)\right],
    \label{current equation}
\end{equation}
where $\mu$ is the carrier mobility, $q_{is}$ and $q_{id}$ are the inversion sheet charge density at source and drain, respectively \cite{Tsormpatzoglou2012FinFET}, which can be obtained by
\begin{equation}
    q_{i s} = \underline{\text{LambertW}}\left(e^{\frac{q\left(V_g-V_t-V_s\right)}{2 k T}}\right),
\end{equation}
\begin{equation}
    q_{i d}=\underline{\text{LambertW}}\left(e^{\frac{q\left(V_g-V_t-V_d\right)}{2 k T}}\right),
\end{equation}
where $V_g$ is the gate voltage, $V_s$ and $V_d$ are the source and drain potential, respectively
\cite{Papathanasiou2012MOSFET}.

\begin{figure*}[t]
    \centering
    \begin{subfigure}[t]{0.7\textwidth}
        \centering
        \includegraphics[width=\textwidth]{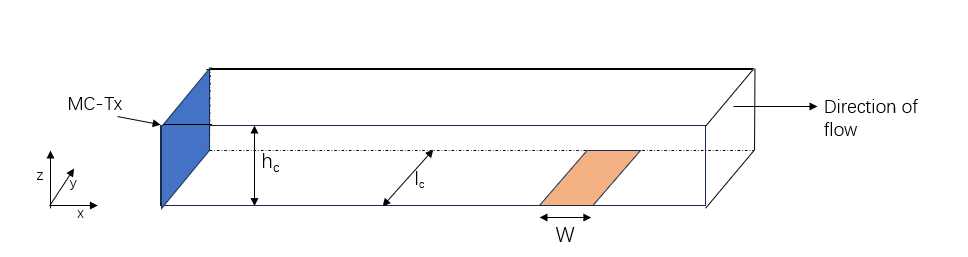}
        \caption{}
        \label{3D views of a microfluidic channel and the locations of transmitter and receiver}
    \end{subfigure}\\
    \begin{subfigure}[t]{0.7\textwidth}
        \centering
        \includegraphics[width=\textwidth]{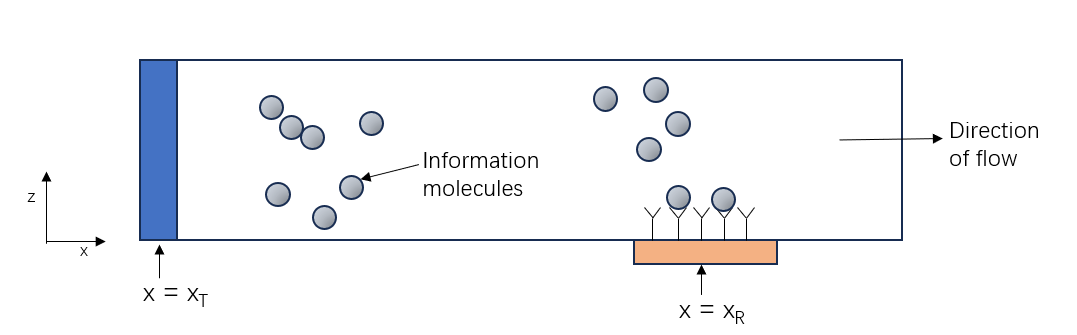}
        \caption{}
        \label{2D views of a microfluidic channel and the locations of transmitter and receiver}
    \end{subfigure}
    \caption{(a) 3D and (b) 2D views of a microfluidic channel and the locations of transmitter and receiver\cite{Kuscu2016SiNWmodeling}}
    \label{(a) 3D and (b) 2D views of a microfluidic channel and the locations of transmitter and receiver}
\end{figure*}

\begin{figure*}[t]
    \centering
    \includegraphics[width = \textwidth]{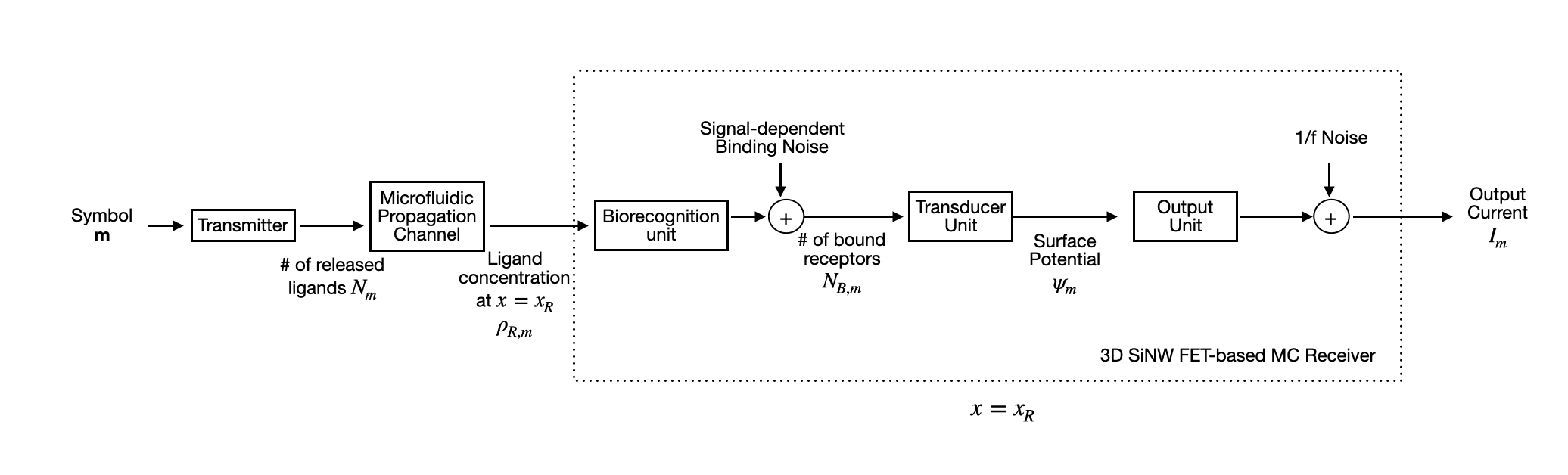}
    \caption{Block diagram of microfluidic MC system with SiNW FET-based MC receiver\cite{Kuscu2016SiNWmodeling}.
    }
    \label{Block diagram of microfluidic MC system with SiNW FET-based MC receiver}
\end{figure*} 

\section{SiNW FinFET Receiver Model}
\label{SiNW FinFET Receiver Model}
In our model, we suggest a time-slotted molecular communication system between a single transmitter-receiver pair that is assumed to be perfectly time-synchronized\cite{Kuscu2016SiNWmodeling}. A rectangular microfluidic channel is used. The planar ligand source transmitter is located at \( x=x_{T} \). The FinFET MC receiver is buried at the bottom at \( x=x_{R} \) in the channel, and we assume that the height of the fin will not significantly change the flow of the fluid. As shown in Figure \ref{(a) 3D and (b) 2D views of a microfluidic channel and the locations of transmitter and receiver}, information is encoded into the concentration of molecules via M-ary concentration shift keying (M-CSK) modulation in this system. Considering that the input alphabet is M = \{0, 1, \ldots, M - 1\}, the transmitter releases \( N_{m} \) molecules at the start of the signaling interval, e.g., at time \( t_{k} = kT_{s} \), where \( T_{s} \) is the slot duration, or the symbol period, in order to communicate the symbol $m \in M$ for the $kth$ time slot. Depending on the concentration of the information molecules, a different current will be generated at the FET-based receiver. In Figure \ref{Block diagram of microfluidic MC system with SiNW FET-based MC receiver}, the block diagram shows the CSK-based MC system with binding noise and $1/f$ noise added. In this work, we only consider 1-bit and 2-bit CSK systems.

\subsection{Molecular Transport in Microfluidic Channel}
The advection-diffusion equation can be used to describe how ligands are transported within the microfluidic channel\cite{Bicen2013MicrofluidicChannels}. Taking into account the effect of Taylor-Aris type dispersion, e.g., inhomogeneous flow fields, solute-wall interactions, and force fields normal to channel walls\cite{Dutta2006MicrofluidicDispersion}, the effective diffusion coefficient \( D \) for a rectangular microfluidic channel can be expressed as

\begin{equation}
    D=\left(1+\frac{8.5 u^2 h_c^2 l_c^2}{210 D_0^2\left(h_c^2+2.4 h_c l_c+l_c^2\right)}\right) D_0,
\end{equation}
where \( D_0 \) is the intrinsic diffusion coefficient, \( l_c \) and \( h_c \) are the cross-sectional channel length and height, respectively. \( u \) is the flow velocity of the fluid\cite{Bicen2013MicrofluidicChannels}. In our propagation model analysis, we neglect Inter-Symbol Interference (ISI). Hence, we only need to consider one signaling interval, e.g., \( k = 0 \) and \( t_k = 0 \). It is possible to express the initial impulse scaled by surface concentration, assuming that ligands are evenly distributed over the channel's cross-section at the release time, the ligand concentration, e.g., $\rho_m(x, t=0)$ can be expressed as 

\begin{equation}
    \rho_m(x, t=0)=\frac{N_m}{A_{c h}} \delta(x),
\end{equation}
where $\delta(x)$ is the Dirac delta function and $A_{c h} = h_c \times l_{ch}$ is the cross-sectional area of the channel. By solving the one-dimensional advection-diffusion equation and consider solution for $t>0$, the ligand concentration profile is given by \cite{Bicen2013MicrofluidicChannels}
\begin{equation}
    \rho_m(x, t)=\frac{N_m / A_{c h}}{\sqrt{4 \pi D t}} \exp \left(-\frac{(x-u t)^2}{4 D t}\right) .
\end{equation}

\subsection{Received Signal}

The peak ligand concentration is attenuated when the ligands diffuse and transported along the channel by the fluid flow. The expected time it takes for the peak concentration to reach the receiver's center position, $x = x_R$, is how we define the propagation delay, i.e., 
\begin{equation}
    t_D=\frac{x_R}{u} ,
\end{equation}
where $u$ is the average flow velocity\cite{Kuscu2016SiNWmodeling}.
Assuming that the transmitter and receiver are perfectly synchronized in time\cite{Kuscu2016SiNWmodeling}, that is the received signal can be detected in the peak value of $ \rho_m(x, t)$ at $t_D$. Therefore, the input signal at the receiver will be
\begin{equation}
    \begin{aligned}
\rho_m\left(x_R, t\right) \approx & \rho_m\left(x_R, t_D\right)=\frac{N_m}{A_{c h} \sqrt{4 \pi D t_D}} \\
& \text { for } t \in\left[t_D-\tau_p / 2, t_D+\tau_p / 2\right],
\label{receiversignal}
\end{aligned}
\end{equation}
where $\tau_p$ is the approximate passage duration of a portion of ligand concentration to be sampled over the receiver surface\cite{Kuscu2016SiNWmodeling}.

\subsection{Biorecognition Block and Binding Noise}
The transport rate, also known as ligand flux to the receiver surface, determines the biorecognition process. For a rectangular cross-section microfluidic channel, transport rate can be approximated as 
\begin{equation}
    k_T=D l_r \times\left\{\begin{array}{l}
\left(0.8075 P_s^{1 / 3}+0.7058 P_s^{-1 / 6}-0.1984 P_s^{-1 / 3}\right), \\
\text { if } P_s>1 \\
\frac{2 \pi}{4.885-\ln \left(P_s\right)}\left(1-\frac{0.09266 P_s}{4.885-\ln \left(P_s\right)}\right),
\text { if } P_s<1
\end{array}\right.
\end{equation}
where $P_s=\left(6 Q w_R^2\right) /\left(D l_{c h} h_{c h}^2\right)$, with $Q = A_{ch}\times u  $ being the volumetric flow rate and $w_R$ being the width of the SiNW \cite{Sheehan2005DetectionLimits}.

In \cite{Berezhkovskii2013ReceptorOccupancy}, it is assumed that there is a stationary ligand concentration fixed to \( \rho_{R,m} \) and reaction equilibrium. Under these conditions, the mean number of bound receptors on the receiver surface, \( \mu_{N_{B, m}}(t) \), can be written as

\begin{equation}
    \mu_{N_{B, m}}=P_{o n \mid m} N_R=\frac{\rho_{R, m}}{\rho_{R, m}+K_D} N_R,
\end{equation}
where \( P_{on \mid m} \) is the probability of finding a single receptor in the ON (bound) state at equilibrium. \( K_D = \frac{k_{-1}}{k_{1}} \) is the dissociation constant, where \( k_{-1} \) and \( k_{1} \) are the effective unbinding and binding rates, respectively. The probability of having \( n \) number of bound receptors can be modeled using a Binomial distribution \cite{Berezhkovskii2013ReceptorOccupancy}, and its variance is expressed as

\begin{equation}
    \sigma_{N_{B, m}}^2=P_{\text {on } \mid m}\left(1-P_{\text {on } \mid m}\right) N_R.
\end{equation}

For the reaction limited case, where $k_T$ have high values, the power spectral density (PSD) of the binding noise can be written as\cite{Kuscu2016SiNWmodeling}

\begin{equation}
    S_{N_{B, m}}(f)=\sigma_{N_{B, m}}^2 \frac{2 \tau_{B, m}}{1+\left(2 \pi f \tau_{B, m}\right)^2},
    \label{bindingnoise}
\end{equation}
where $\tau_{B, m}=\frac{1}{k_1 \rho_{R, m}+k_{-1}}+\frac{k_1\left(k_1 \rho_{R, m}+N_R k_{-1}\right)}{k_T\left(k_1 \rho_{R, m}+k_{-1}\right)^2} $ is the relaxation time of transport-influenced ligand-receptor binding reaction. $\tau_p$, or the amount of time over which the ligand concentration is taken to be constant, limits the equilibrium assumption for receptors. When receptors are exposed to a steady-state ligand concentration for a duration that satisfies $\tau_p \geq 5 \tau_{B, m}$, they are considered to be in an equilibrium state since the relaxation time $\tau_{B, m}$ dictates the time to reach equilibrium\cite{Kuscu2016SiNWmodeling}.

\subsection{Transducer Block}
We assume the charge accumulated due to transduction of the ligands on the top and sides surface of the gate will add up equally. The surface potential would 
then be
\begin{equation}
    \Psi_m=\frac{Q_m}{C_{e q}},
\label{psi_m}
\end{equation}
where $Q_m = N_{B, m} q_{e f f} N_{e^{-}} $ is the charge generated on the surface and $C_{eq}$ is the equivalent capacitance of the transducer\cite{Curreli2008NanowireFETs}. Here, $q_{eff}$ is the mean effective charge due to Debye screening\cite{Stern2007DebyeScreening}.

Since the greater the distance between the ligand electron and the transducer, the lower the mean effective charge of the free ligand electron, the mean effective charge can be further expressed as 
\begin{equation}
    q_{e f f}=q \times \exp \left(-r / \lambda_D\right),
\end{equation}
where $q$ is the elementary charge, $\lambda_D$ is the Debye length and $r$ is the average distance of ligand electrons in the bound state to the transducer’s surface\cite{Rajan2013NanowireBiosensors}.

We assume the average distance $r$ is equal to the average surface receptor length\cite{Kuscu2016SiNWmodeling}. The Debye length $\lambda_D$ can be expressed as 
\begin{equation}
    \lambda_D=\sqrt{\left(\epsilon_M k_B T\right) /\left(2 N_A q^2 c_{i o n}\right)},
\end{equation}
where $\epsilon_M$ is the dielectric permittivity of the medium, $k_B$ is the Boltzmann’s constant, $T$ is the temperature, $N_A$ is Avogadro’s number, and $c_{ion}$ is the ionic concentration of the medium \cite{Rajan2013NanowireBiosensors}.

The equivalent capacitance described in (\ref{psi_m}), $C_{eq}$ is given in \cite{Kuscu2016SiNWmodeling}
\begin{equation}
    C_{e q}=\left(\frac{1}{C_{O X}}+\frac{1}{C_{N W}}\right)^{-1}+C_{D L},
\end{equation}
Where oxide layer capacitance $C_{O X}=\left(\epsilon_{O X} / t_{O X}\right) w_{eff} l_R$\cite{Shoorideh2014NanoscaleFETBiosensors}, diffusion layer capacitance $ C_{DL} =\left(\epsilon_M / \lambda_D\right) w_{eff} l_R$ and the double layer capacitance emerged in the NW channel $C_{N W}=\left(\epsilon_{S i} / \lambda_{NW}\right) w_{eff} l_R$ \cite{Gao2010NanowireFETBiosensors}. $t_{ox}$ and $\epsilon_{ox}$ are the thickness and the permittivity of the oxide respectively. $\lambda_{NW}$ is the thickness of the double layer created in the inner surface of the $NW$ where $\lambda_{N W}=\sqrt{\left(\epsilon_{S i} k_B T\right) /\left(p q^2\right)}$. Here, $p$ is the hole density.

\subsection{Output Block and $1/ f$ Noise}
The potential generated due to ligand charges induced at the SiNW-oxide layer produces a variation in the current flow through the channel. For the simplicity of comparing the FinFET-based device to previous planar analysis, we assume that the magnitude of the current, as produced by (\ref{current equation}), is still valid for a p-type FinFET by setting \( \mu = \mu_p \), where \( \mu_p \) is the hole mobility. The device will operate in the linear region, i.e., 

\begin{equation}
    V_{S G}>\left|V_t\right| ; \quad V_{S D} \leq V_{S G}-\left|V_t\right|
\end{equation}
where $V_{SG}$ is the source to gate voltage, $V_{SD}$ is the source to drain voltage.

The transconductance, which is the partial derivative of the source-drain current with respect to source-gate voltage, is given as
\begin{equation}
    g_{F E T}=\frac{\partial I_{DS }}{\partial V_{GS}}.
\end{equation}
The mean of the generated output current is given by
\begin{equation}
    \mu_{I_m}=g_{F E T} \Psi_L N_R\left(1+\frac{K_D A_{c h}}{N_m} \sqrt{\frac{4 \pi D x_R}{u}}\right)^{-1},
    \label{expectedoutput}
\end{equation}
where surface potential created by a single ligand $\Psi_L=\left(q_{e f f} \times N_e^{-}\right) / C_{e q}$ and $N_R$ is the number of receptors on the surface\cite{Kuscu2016SiNWmodeling}.

The bioFET-based MC receiver suffers from flicker noise during low-frequency operation, similar to other transistor devices. The flicker noise for FET type devices is accurately described by the correlated carrier number and mobility fluctuation model, which we use in this paper. We attribute the source of the flicker noise to the random generation and recombination of charge carriers as a result of defects and traps in the SiNW channel brought about by imperfect fabrication\cite{Sze2007SemiconductorDevices}. The resulting output current-referred noise PSD at triode region is \cite{Rajan2010TemperatureDependence}
\begin{equation}
    S_{I_m^F}(f)=S_{V, F B}(f) g_{F E T}^2\left[1+\alpha_s \mu_n C_{O X}\left(V_{SG}-\left|V_{t}\right|\right)\right]^2,
    \label{flickernoise}
\end{equation}
where the PSD of the flatband-voltage noise $S_{V, F B}(f)$ is given as 
\begin{equation}
    S_{V, F B}(f)=\frac{\lambda k_B T q^2 N_{o t}}{w_R l_R C_{O X}^2|f|},
\end{equation}
where $\lambda$ is the characteristic tunneling distance, $N_{ot}$ is the oxide trap density, i.e., impurity concentration, of the SiNW channel\cite{Rajan2010TemperatureDependence}.

\subsection{Overall Noise PSD}
The overall PSD of the output current referred noise can be given as
\begin{equation}
    S_{I_m}(f)=S_{I_m^B}(f)+S_{I_m^F}(f),
\end{equation}
Where additive binding noise $S_{I_m^B} =S_{N_{B, m}}(f) \times \Psi_L^2 \times g_{F E T}^2$. Therefore, the SNR at the output will be
\begin{equation}
    S N R_{\text {out }, m}=\frac{\mu_{I_m}^2}{\sigma_{I_m}^2},
\end{equation}
where $\sigma_{I_m}^2$ is the output current variance which can be computed as follows
\begin{equation}
    \sigma_{I_m}^2=\int_{-\infty}^{\infty} S_{I_m}(f) d f.
\end{equation}

\begin{table}[h!]
  \centering
  \caption{\textit{Default Values of Simulation Parameters}}
  \label{tab:simulation_parameters}
  \begin{tabular}{ll}
    \hline
    \textbf{\textit{Parameter}} & \textbf{\textit{Value}} \\
    \hline
    \textit{Flatband voltage\((V_{fb}))\)}&\textit{\(-0.4762 \, V\)} \\
    \textit{SiNW height\((t_s))\)}&\textit{\(5 \times 10^{-8} \, m\)} \\
    \textit{Microfluidic channel height \((h_{ch})\)} & \textit{\(3 \, \mu m\)} \\
    \textit{Microfluidic channel width \((l_{ch})\)} & \textit{\(15 \, \mu m\)} \\
    \textit{Number of transmitted ligands for symbol \(m (N_m)\)} & \textit{\(5 \times 10^5\)} \\
    \textit{Max number of ligands transmitter can release \((K)\)} & \textit{\(4 \times 10^6\)} \\
    \textit{Transmitter-receiver distance \((d)\)} & \textit{\(1 \, mm\)} \\
    \textit{Average flow velocity \((u)\)} & \textit{\(10 \, \mu m/s\)} \\
    \textit{Diffusion coefficient of ligands \((D_0)\)} & \textit{\(2 \times 10^{-10} \, m^2/s\)} \\
    \textit{Binding rate \((k_1)\)} & \textit{\(2 \times 10^{-19} \, m^3/s\)} \\
    \textit{Unbinding rate \((k_{-1})\)} & \textit{\(20 \, s^{-1}\)} \\
    \textit{Average number of electrons in a ligand \((N_e^-)\)} & \textit{\(3\)} \\
    \textit{Width of the SiNW \((W)\)} & \textit{\(10 \pi \, nm\)} \\
    \textit{Concentration of receptors on the surface \((\rho_{SR})\)} & \textit{\(4 \times 10^{16} \, m^{-2}\)} \\
    \textit{Length of a surface receptor \((l_{SR})\)} & \textit{\(2 \, nm\)} \\
    \textit{Temperature \((T)\)} & \textit{\(300K\)} \\
    \textit{Relative permittivity of oxide layer \((\varepsilon_{ox} / \varepsilon_0)\)} & \textit{\(3.9\)} \\
    \textit{Relative permittivity of SiNW \((\varepsilon_{NW} / \varepsilon_0)\)} & \textit{\(11.68\)} \\
    \textit{Relative permittivity of medium \((\varepsilon_{r} / \varepsilon_0)\)} & \textit{\(78\)} \\
    \textit{Ionic strength of electrolyte medium \((c_{ion})\)} & \textit{\(30 \, mol/m^3\)} \\
    \textit{Source-drain voltage \((V_{SD})\)} & \textit{\(0.1 \, V\)} \\
    \textit{Overhead voltage \((V_{SD}- V_{t})\)} & \textit{\(0.4 \, V\)} \\
    \textit{Hole density in SiNW \((p)\)} & \textit{\(10^{18} \, cm^{-3}\)} \\
    \textit{Tunneling distance \((\lambda)\)} & \textit{\(0.05 \, nm\)} \\
    \textit{Thickness of oxide layer \((t_{ox})\)} & \textit{\(2 \, nm\)} \\
    \textit{Oxide trap density \((N_{ot})\)} & \textit{\(10^{16} \, eV^{-1}cm^{-3}\)} \\
    \textit{Effective mobility of hole carriers \((\mu_p)\)} & \textit{\(500 \, cm^2/Vs\)} \\
    \textit{Coulomb scattering coefficient \((\alpha_s)\)} & \textit{\(1.9 \times 10^{-4} \, Vs/C\)} \\
    \hline
  \end{tabular}
  \label{Default Values}
\end{table}

\section{Performance Analysis}
\label{Performance Analysis}
In this section, we present numerical results derived from our model, using various settings to demonstrate the efficacy of concentration shift keying (CSK) with a 3D MC receiver. Table \ref{Default Values} outlines the default settings for configurable parameters used in our studies. These settings are based on those used in a planar MC receiver, as referenced in \cite{Kuscu2016SiNWmodeling}. The original work did not include the threshold voltage in its analysis; therefore, we use the same overhead voltage ($V_{SG}-V_t$) to ensure a fair comparison. We have chosen a Fin height of $5 \times 10^{-8}m$, assuming it is significantly smaller than the channel height and thus does not affect the fluid flow.

\subsection{Receiver Response and Noise Power}

\begin{figure}[t]
    \centering
    \includegraphics[width = 0.40\textwidth]{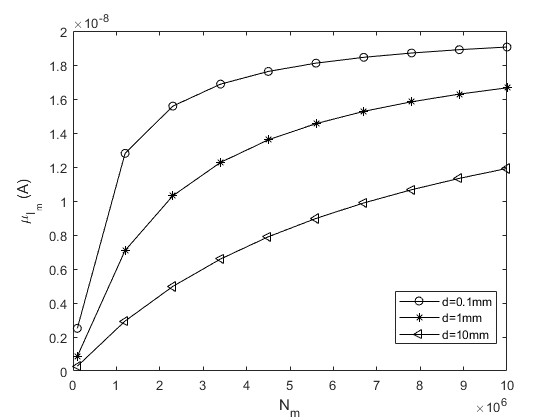}
    \caption{Expected output current $\mu_{I_m}$ as a function of number of ligands $N_m$ released by transmitter in 3D receiver}
    \label{Expected output current}
\end{figure}
\begin{figure}[t]
    \centering
    \includegraphics[width = 0.40\textwidth]{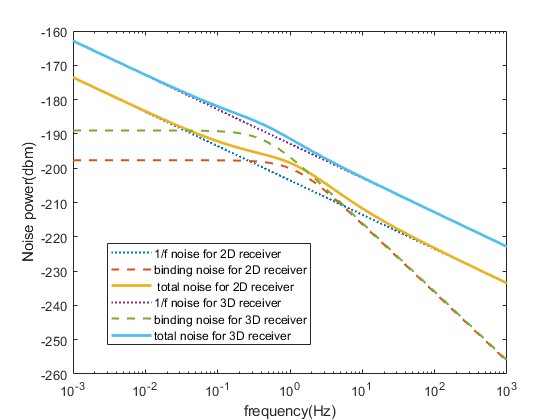}
    \caption{PSD of noise on SiNW based receiver in 2D and 3D cases}
    \label{PSD of noise on SiNW based receiver in 2D and 3D cases}
\end{figure}

\subsubsection{\textit{Receiver Response}}
In Figure \ref{Expected output current}, the expected current response at the receiver due to the variation in the number of ligands released by the transmitter, denoted as \( N_m \), is shown. We also investigate the impact of different transmission distances on this response.

The performance observed is similar to the 2D cases referred to in \cite{Kuscu2016SiNWmodeling}. As the number of transmitted ligands increases, a higher output current is generated. However, beyond a certain threshold, owing to the saturation of surface receptors, the device begins to lose its sensitivity to variations in ligand concentration.

Transmission distance is another crucial factor affecting the performance of the output current. The concentration attenuates by a factor of \( \sqrt{d} \) with an increase in transmission distance, as shown in (\ref{receiversignal}). The shorter the transmission length, the more sensitive the output current is to the number of transmitted information molecules. However, this also leads to quicker saturation. For instance, in the 10mm transmission distance case, the expected output current does not saturate even at \( N_m = 10^7 \), but the sensitivity is reduced, potentially leading to detection issues.

The results obtained, when compared to 2D cases, show a significantly higher expected output level. This is attributed to the new proposed device having a much larger area of receptors, allowing for more charges to accumulate on its surface. Additionally, the FinFET structure, with its better-controlled channel, results in a higher transconductance, further improving the expected output current, as seen in (\ref{expectedoutput})).

\subsubsection{\textit{Noise Power}}
The PSD of noise on SiNW based receiver in 2D and 3D cases are plotted in Figure (\ref{PSD of noise on SiNW based receiver in 2D and 3D cases}). The individual PSDs of binding and flicker noises are plotted as well, respectively. At low frequencies, $1/f$ noise dominates for both cases since binding noise has a flat power density at frequencies below \( \frac{1}{\tau_B} \)\cite{Kuscu2016SiNWmodeling}. At high frequencies (\( f > 10Hz \)), the binding noise attenuates more than flicker noise, hence, both 2D and 3D models are dominated by flicker noise. Around its cut-off frequency, binding noise starts to dominate in the 2D case, whereas in the 3D case, both binding noise and flicker noise contribute to the total noise almost equally. The cut-off frequency is reduced due to the 3D structure since it has a higher number of surface receptors (see (\ref{bindingnoise})). The flicker noise is higher in 3D cases. Since the flicker noise PSD is proportional to \( g_{FET}^2 \) (see (\ref{flickernoise})), and due to its larger effective width, the 3D structure has a higher transconductance.

\subsection{SNR analysis}
\begin{figure*}[t]  
    \centering
    \begin{subfigure}[b]{0.24\textwidth}
        \centering
        \includegraphics[width=\textwidth]{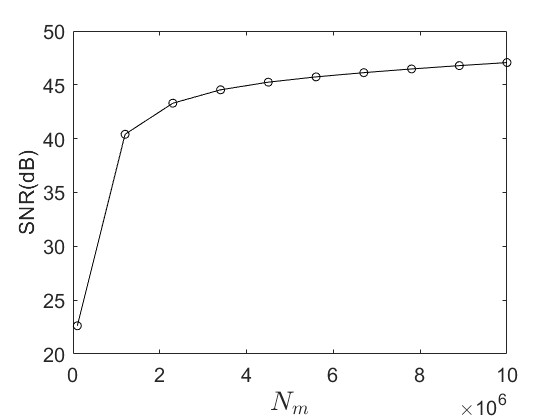}
        \caption{}
        \label{SNRNm}
    \end{subfigure}
    \hfill
    \begin{subfigure}[b]{0.24\textwidth}
        \centering
        \includegraphics[width=\textwidth]{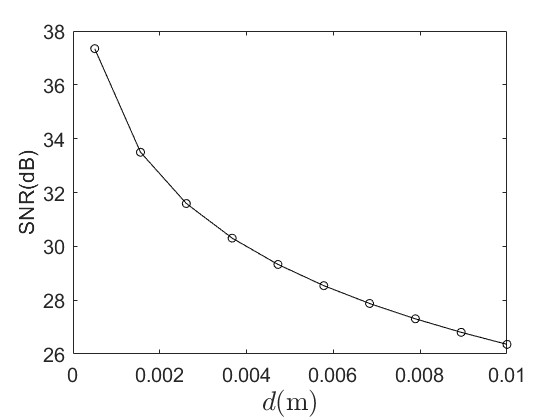}
        \caption{}
        \label{SNRd}
    \end{subfigure}
    \hfill
    \begin{subfigure}[b]{0.24\textwidth}
        \centering
        \includegraphics[width=\textwidth]{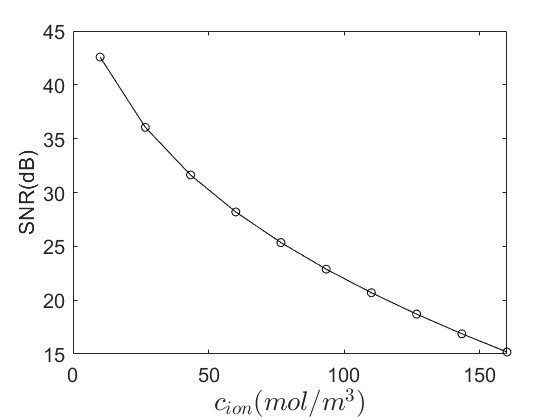}
        \caption{}
        \label{SNRcion}
    \end{subfigure}
    \hfill
    \begin{subfigure}[b]{0.24\textwidth}
        \centering
        \includegraphics[width=\textwidth]{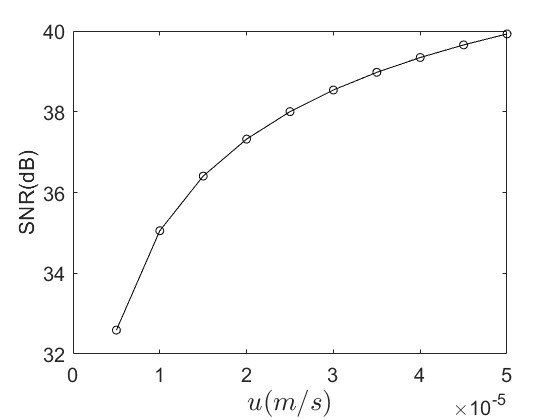}
        \caption{}
        \label{SNRu}
    \end{subfigure}
    
    \caption{Effect of the communication system parameters on the SNR at the electrical output of the  FinFET-based receiver. SNR as a function of (a) number of transmitted ligands $N_m$, (b) transmitter-receiver distance d, (c) ion concentration $c_{ion}$ of the electrolyte medium, (d) average flow velocity u inside the microfluidic channel.}
    \label{fig:combined}
\end{figure*}

\begin{figure*}[t]  
    \centering
    
    \begin{subfigure}[b]{0.24\textwidth}
        \centering
        \includegraphics[width=\textwidth]{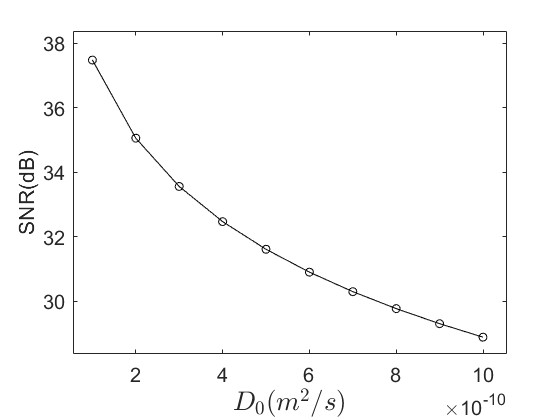}
        \caption{}
        \label{SNRD0}
    \end{subfigure}
    \hfill
    \begin{subfigure}[b]{0.24\textwidth}
        \centering
        \includegraphics[width=\textwidth]{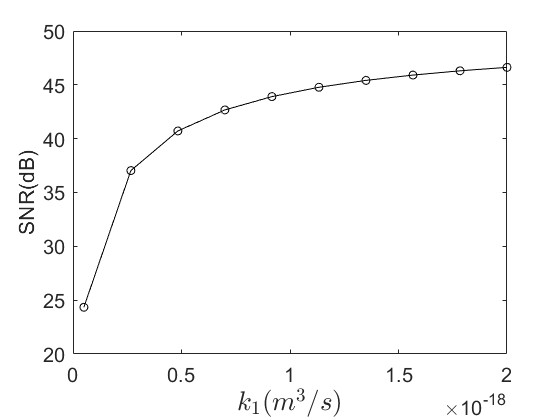}
        \caption{}
        \label{SNRk1}
    \end{subfigure}
    \hfill
    \begin{subfigure}[b]{0.24\textwidth}
        \centering
        \includegraphics[width=\textwidth]{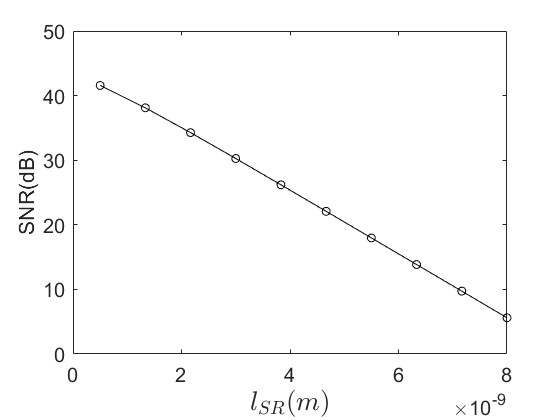}
        \caption{}
        \label{SNRISR}
    \end{subfigure}
    \hfill
    \begin{subfigure}[b]{0.24\textwidth}
        \centering
        \includegraphics[width=\textwidth]{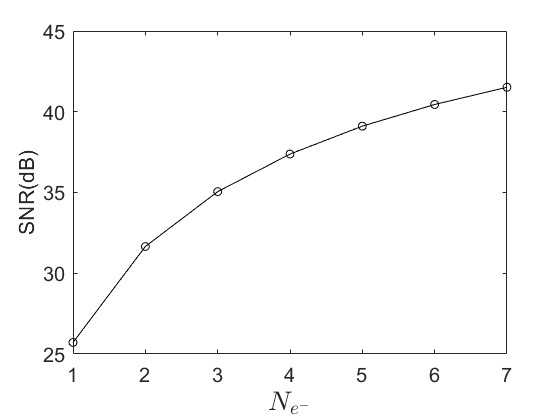}
        \caption{}
        \label{SNRNe}
    \end{subfigure}
    
    \caption{Effect of the molecular parameters on the SNR at the electrical output of the FinFET-based receiver. SNR as a function of (a) intrinsic diffusion coefficient of ligands $D_0$, (b) intrinsic binding rate of ligands $k_1$, (c) surface receptor length $l_{SR}$, (d) number of free electrons per ligand molecule $N_e^{-}$}
    \label{}
\end{figure*}

\begin{figure*}[t]  
    \centering
    
    \begin{subfigure}[b]{0.24\textwidth}
        \centering
        \includegraphics[width=\textwidth]{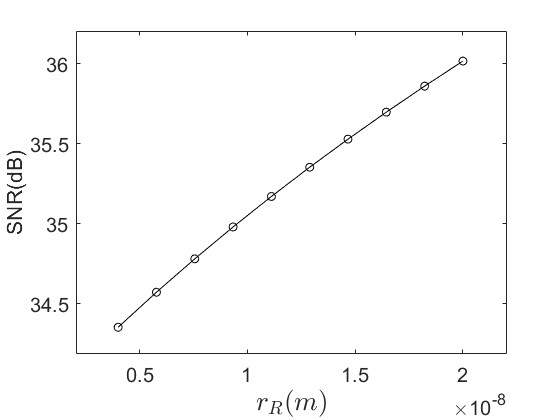}
        \caption{}
        \label{SNRr_R}
    \end{subfigure}
    \hfill
    \begin{subfigure}[b]{0.24\textwidth}
        \centering
        \includegraphics[width=\textwidth]{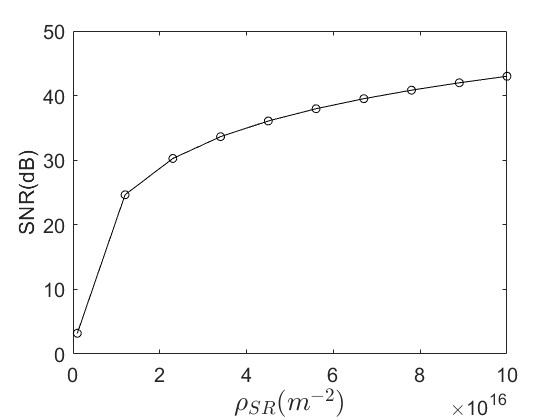}
        \caption{}
        \label{SNRrhosr}
    \end{subfigure}
    \hfill
    \begin{subfigure}[b]{0.24\textwidth}
        \centering
        \includegraphics[width=\textwidth]{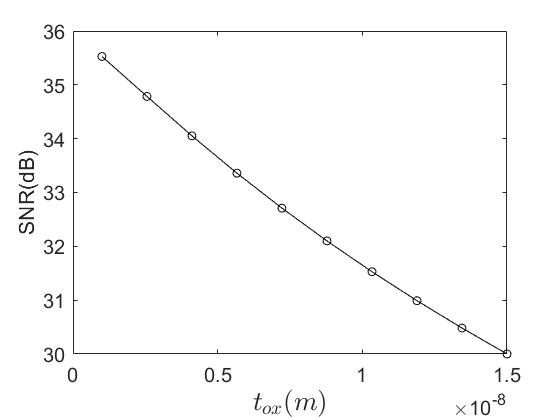}
        \caption{}
        \label{SNRtox}
    \end{subfigure}
    \hfill
    \begin{subfigure}[b]{0.24\textwidth}
        \centering
        \includegraphics[width=\textwidth]{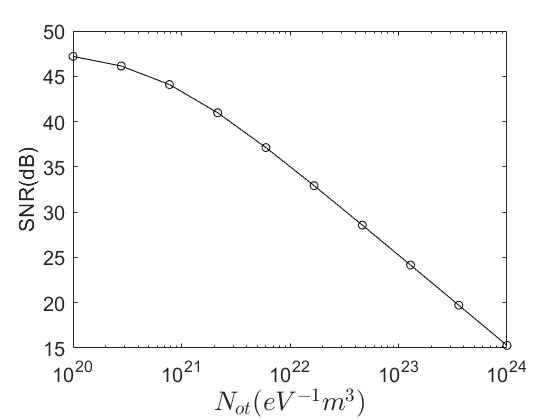}
        \caption{}
        \label{SNRNot}
    \end{subfigure}
    
    \caption{Effect of the receiver design parameters on the SNR at the electrical output of the FinFET-based receiver. SNR as a function of (a) SiNW radius $r_R$, (b) surface receptor concentration $\rho_{SR}$, (c) $SiO_2$ layer thickness $r_{ox}$, and (d) oxide trap density $N_{ot}$ in SiNW.}
    \label{}
\end{figure*}
In this section, the SNR at the output of our proposed FinFET-based receiver is investigated under varying system parameters. These parameters are further divided into three groups: (i) communication system related parameters, (ii) information molecules and receptors related parameters, and (iii) receiver related parameters. \( r_R \) is defined as the SiNW radius, which can be approximated as \( \frac{W_{fin}}{\pi} \).

\subsubsection{\textit{Effect of Communication System-related Parameters}} 
SNR is first investigated against various numbers of ligands released at the transmitter in Figure \ref{SNRNm}. From the graph plotted, it is observed that the more ligands transmitted, the higher the SNR that can be achieved. The SNR starts to saturate at 45dB when the surface receptors are exposed to a high concentration of ligands. Compared to the 2D receiver, the saturation SNR is 5dB higher because of a larger surface area to accommodate receptors, which can produce a higher expected current. This outweighs the effect of an increase in noise. In Figure \ref{SNRd}, for a fixed number of transmitted ligands, the output SNR decreases with an increase in transmission distance. This is because the concentration of information molecules is attenuated by \( \sqrt{d} \) as the distance increases.

Figure \ref{SNRcion} shows how the ionic strength of the fluid affects the output SNR. The larger the ionic strength of the fluid, the lower the Debye length. As a result, the effective charge will be smaller, leading to a reduced potential generated at the gate. This results in a smaller output current and, consequently, a lower SNR.

The effect of fluid flow \( u \) on the receiver SNR is shown in Figure \ref{SNRu}. An increase in \( u \) means that the information molecules move faster towards the receiver, resulting in less attenuation. Therefore, the SNR improves before reaching saturation.

For the FinFET-based receiver, the SNR exhibits a similar trend compared to the 2D receiver, with overall noise performance improved against communication parameters.

\subsubsection{\textit{Effect of Information Molecules and Receptors-related Parameters}}
The diffusion coefficient \( D_0 \) has a significant effect on the output SNR, as shown in Figure \ref{SNRD0}. A higher \( D_0 \) leads to a higher effective diffusion coefficient \( D \). Since an increase in \( D \) attenuates the concentration of molecules at the receiver by \( \sqrt{D} \), a higher \( D_0 \) results in a lower SNR at the output. 

The intrinsic binding rate \( k_1 \) measures the rate at which ligands can combine with the receptors as they flow over the receiver. A higher \( k_1 \) leads to a lower dissociation constant \( K \), which increases the mean number of bound receptors, thus increasing the expected output current and SNR, as shown in Figure \ref{SNRk1}. 

The receptor length is an important receiver characteristic. A higher receptor length reduces the effective charge, hence reducing the output level. An almost linear reduction of SNR in dB can be observed from Figure \ref{SNRISR}. 

Finally, the effect of the number of free charges per ligand is investigated and shown in Figure \ref{SNRNe}. Highly charged ligands generate more potential at the gate, leading to a higher output current, which results in a better SNR.

\subsubsection{\textit{Effect of Receiver-related Parameter}}
The size of SiNW also affects the performance of the receiver, as shown in Figure \ref{SNRr_R}. \( r_R \) is directly related to the width of the FET-based receiver. A larger width implies a larger area of receptors, hence more ligands can be accommodated. Moreover, the capacitance of the oxide layer and diffusion layer are all related to this parameter. The increased SNR shows the same trend compared to 2D cases, whereas the increase in SNR performance due to the increase in \( r_R \) is less pronounced compared to the 2D receiver. This is because the 3D structure has a larger effective width compared to the 2D one, making the percentage increase in \( W \) smaller. Figure \ref{SNRrhosr} shows how the density of the receptor affects the performance of the output SNR. A higher number of receptors will potentially produce a higher fluctuation in gate voltage, hence, producing a better SNR. At a high level of surface receptor, the SNR begins to saturate because the transmitted signal is less than the number of surface receptors, indicating that an increase in surface receptors may not necessarily lead to a higher number of bindings. 

The effect of the thickness of the oxide layer \( t_{ox} \) is shown in Figure \ref{SNRtox}. A thicker \( t_{ox} \) leads to a decrease in \( C_{OX} \), which reduces the transconductance of the FET-based receiver. Therefore, less gate voltage fluctuation can be transduced to the current signal, leading to a lower SNR. In the FinFET-based receiver, the trap density has a negative impact on the output SNR, as shown in Figure \ref{SNRNot}. This is because a larger \( N_{ot} \) leads to a higher flatband-voltage noise. A higher flatband-voltage noise causes a higher flicker noise PSD, which reduces the SNR.

\subsection{SEP Analysis in physiological solutions}
SNR analysis shows superior performance for the proposed 3D FinFET-based receiver, which utilizes vertical space. However, there are still some design parameters that need to be considered. The gate voltage applied, after considering the threshold voltage, is around -0.9V, which seems to be too large for a device ultimately aimed at harvesting energy from the surroundings. Moreover, the ionic strength used in previous analysis is too small if we aim at organoid communication. For example, Bovine serum and saline, often used as dialysate, have an ionic strength around \( 150 \, \text{mol/m}^3 \) \cite{Okada1990DialysisMembrane}. A higher fin is used in the following analysis with \( t_s = 5 \times 10^{-7} \, \text{m} \).

SEP based on maximum likelihood (ML) detection is given as 
\begin{equation}
    \begin{aligned}
P_e = \frac{1}{2 M} & \left[\operatorname{erfc}\left(\frac{\lambda_1 - \mu_{I_0}}{\sigma_{I_0} \sqrt{2}}\right) + \operatorname{erfc}\left(\frac{\mu_{I_{M-1}} - \lambda_{M-1}}{\sigma_{I_{M-1}} \sqrt{2}}\right)\right] \\
+ & \sum_{m=1}^{M-2} \left(\operatorname{erfc}\left(\frac{\mu_{I_m} - \lambda_m}{\sigma_{I_m} \sqrt{2}}\right) + \operatorname{erfc}\left(\frac{\lambda_{m+1} - \mu_{I_m}}{\sigma_{I_m} \sqrt{2}}\right)\right),
\end{aligned}
\end{equation}
where \( \mu_{I_m} \) is the mean output current and \( \sigma_{I_m} \) is the output current variance for symbol m, i.e., \( m = 0, \ldots, M-1 \), \(\operatorname{erfc}(x)\) is the complementary error function, and \( \lambda_m \) is the decision threshold based on ML\cite{Kuscu2016SiNWmodeling}.

\begin{figure*}[t]  
    \centering
    \begin{subfigure}[b]{0.32\textwidth}
        \centering
        \includegraphics[width=\textwidth]{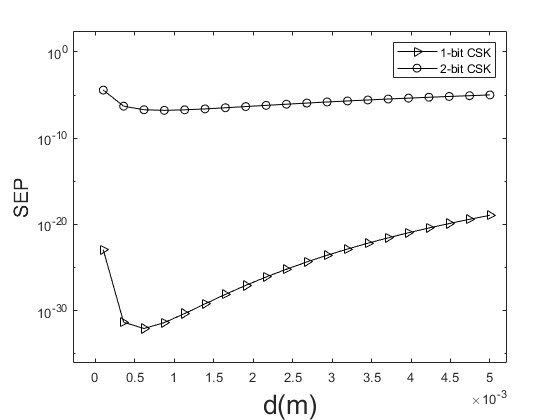}
        \caption{}
        \label{SEPd}
    \end{subfigure}
    \hfill
    \begin{subfigure}[b]{0.32\textwidth}
        \centering
        \includegraphics[width=\textwidth]{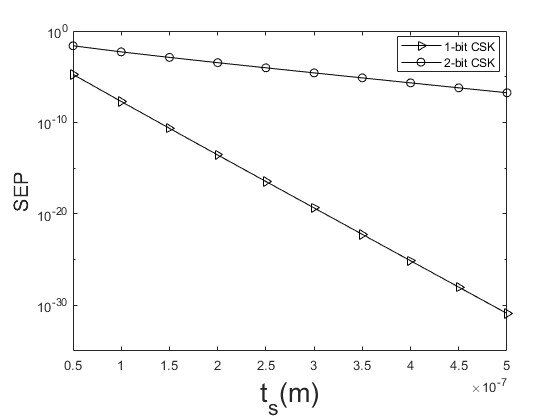}
        \caption{}
        \label{SEPt_s}
    \end{subfigure}
    \hfill
    \begin{subfigure}[b]{0.32\textwidth}
        \centering
        \includegraphics[width=\textwidth]{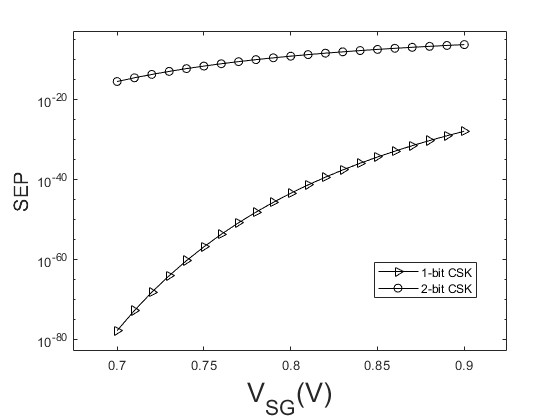}
        \caption{}
        \label{SEPVgs}
    \end{subfigure}
    \caption{Symbol error probability (SEP) for 1-bit and 2-bit modulation. SEP as a function of (a) transmission distance d, (b) height of the SiNW $t_s$, (c) source-drain voltage $V_{SD}$}
\end{figure*}

\begin{figure*}[t]  
    \centering
    \begin{subfigure}[b]{0.32\textwidth}
        \centering
        \includegraphics[width=\textwidth]{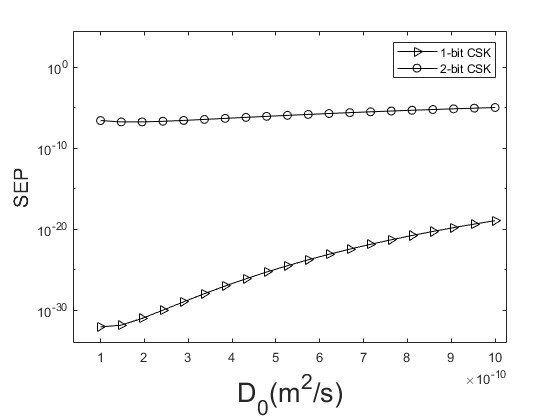}
        \caption{}
        \label{SEPd0}
    \end{subfigure}
    \hfill
    \begin{subfigure}[b]{0.32\textwidth}
        \centering
        \includegraphics[width=\textwidth]{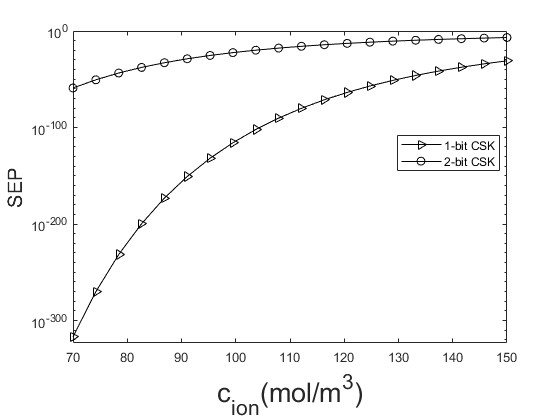}
        \caption{}
        \label{SEPcion}
    \end{subfigure}
    \hfill
    \begin{subfigure}[b]{0.32\textwidth}
        \centering
        \includegraphics[width=\textwidth]{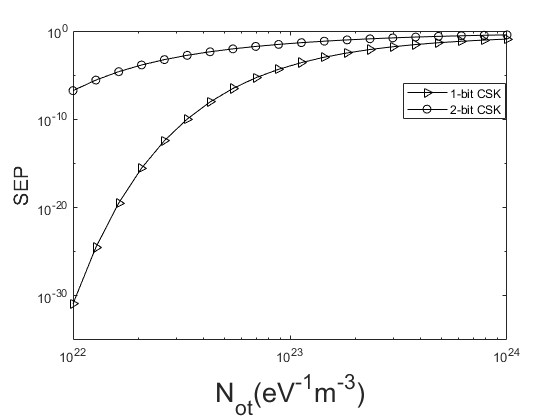}   
        \caption{}
        \label{SEPNOT}
    \end{subfigure}
    \caption{Symbol error probability (SEP) for 1-bit and 2-bit modulation. SEP as a function of (a) intrinsic diffusion coefficient $D_0$, (b) ion concentration of the electrolyte $c_{ion}$, (c) oxide trap density $N_{ot}$}
\end{figure*}

In the following analysis, 1-bit and 2-bit Concentration Shift Keying (CSK) will be taken into account. Figure \ref{SEPd} shows how the SEP changes as a function of transmission distance. It is evident that at short transmission distances, the receiver tends to saturate, and at longer transmission distances, the concentration of information molecules attenuates, hence resulting in a high SEP. Therefore, it is clear that certain optimizations can be performed on the transmission distance to achieve the optimal transmission distance. Compared to 1-bit CSK, 2-bit CSK exhibits a higher SEP but is less sensitive to changes in transmission distance.

One of the key parameters in our proposed 3D model is the height of the fin, which is almost the same magnitude as the height of SiNW \( t_s \), as shown in Figure \ref{SEPt_s}. Devices with a thicker \( t_s \) tend to have a higher number of surface receptors, hence creating a higher expected output, with each symbol then being further apart. Therefore, as \( t_s \) increases, SEP decreases. Note that as \( t_s \) increases, it will eventually affect the fluid flow and ligand propagation. Hence, a higher channel height should be used in the analysis, or a new received signal equation should be derived.

\( V_{SD} \) is an important parameter if we consider the power consumption of the device. The relationship between \( V_{SD} \) and SEP is shown in Figure \ref{SEPVgs}. A larger magnitude of \( V_{SD} \) causes a higher level of flicker noise, which reduces the SEP level. Therefore, as long as the device is operating in the linear region, the lower the \( V_{SD} \), the better the SEP performance.

The intrinsic diffusion coefficient \( D_0 \) is a crucial parameter in physiological solutions. As shown in Figure \ref{SEPd0}, a higher \( D_0 \) leads to a higher SEP. This is due to the fact that a higher diffusion coefficient will attenuate the signal received. To cope with different ionic strengths in different physiological solutions, SEP is found against different \( c_{ion} \). A higher ionic strength leads to a low Debye length, hence a lower effective charge. This makes it difficult for the system to distinguish different output levels, leading to a high SEP. The oxide trap density \( N_{ot} \) indicates the impurity within the channel. A lower \( N_{ot} \) means a clearer channel and a lower flicker noise. Hence, as \( N_{ot} \) increases, a higher SEP is observed, as shown in Figure \ref{SEPNOT}. In general, the proposed 3D receiver demonstrates the capability to operate effectively in more challenging environments compared to 2D counterparts, while still maintaining an acceptable SEP.

\section{Conclusions}
\label{Conclusion}
The implementation of a 3D FinFET-based molecular communication receiver for organoid communication was investigated for the first time in the literature, where both the top and side gates are covered by receptors. The model takes into account the height of the fin and the threshold voltage. A rectangular microfluidic channel is considered, and both biological binding noise as well as flicker noise are included in the analysis. The results based on Signal-to-Noise Ratio (SNR) and Symbol Error Probability (SEP) for the proposed 3D receiver have been compared to those of the 2D receiver, demonstrating better noise performance and good SEP under physiological conditions.

\bibliographystyle{IEEEtran}
\bibliography{references}

\vfill

\end{document}